\documentclass[prl,twocolumn,showpacs,amsmath,amssymb]{revtex4}
\usepackage{graphicx}
\usepackage{dcolumn}
\usepackage{bm}

\begin{document}

\title{\boldmath
  Measurements of the continuum $R_{\rm uds}$ and $R$ values in $e^+e^-$
annihilation in the energy region between 3.650 and 3.872 GeV}
 \author{
\begin{small}
M.~Ablikim$^{1}$,      J.~Z.~Bai$^{1}$,            Y.~Ban$^{11}$,
J.~G.~Bian$^{1}$,      X.~Cai$^{1}$,               H.~F.~Chen$^{15}$,
H.~S.~Chen$^{1}$,      H.~X.~Chen$^{1}$,           J.~C.~Chen$^{1}$,
Jin~Chen$^{1}$
% ...
\end{small}
\\(BES Collaboration)\\
}
\affiliation{ 
\begin{minipage}{145mm}
$^{1}$ Institute of High Energy Physics, Beijing 100049, People's Republic
of China\\
% ...
$^{11}$ Peking University, Beijing 100871, People's Republic of China\\
% ...
$^{15}$ University of Science and Technology of China, Hefei 230026,
People's Republic of China\\
% ...
\end{minipage}
}
\begin{abstract}
We report measurents of the continuum $R_{\rm uds}$ 
near the center-of-mass energy of 3.70 GeV, 
the $R_{{\rm uds(c)}+\psi(3770)}(s)$
and the $R_{\rm had}(s)$ values 
in $e^+e^-$ annihilation at 68 energy points in the energy region 
between 3.650 and 3.872 GeV with the BES-II detector at the BEPC Collodier.
We obtain the $R_{\rm uds}$ 
for the continuum light hadron (containing u, d and s quarks)
production near the $D\bar D$ threshold
to be $R_{\rm uds}=2.141 \pm 0.025 \pm 0.085$. 
\end{abstract}

\maketitle
     Precision measurement of the cross section $\sigma^0_{\rm had}(s)$ for
single photon exchange process in $e^+e^-$ annihilation at the c.m.
(center-of-mass) energy $\sqrt{s}$, which does not
include the initial state radiative and vacuum polarization effects,
are important for testing
the vailidity of the perturbative QCD (pQCD) calculation on the cross section
and in calculation of the photon vacuum polarization.
This cross section is defined as the lowest order cross section for inclusive
hadron production in the paper.
It is often represented by a $R(s)$ ratio 
\begin{equation}
R(s)=\frac{\sigma^{0}_{\rm had}(s)}{\sigma^{0}_{\mu^+\mu^-}(s)},
\end{equation}  
where
$\sigma^{0}_{\mu^+\mu^-}(s)={4\pi \alpha^2(0)}/{3s}$   
is the lowest order QED cross section for muon pair production.
The $\sigma^{0}_{\rm had}(s)$   
still includes the final state photon radiation and
all QCD corrections.
It can be extracted from the observed cross section 
$\sigma^{\rm obs}_{\rm had}(s)$ for inclusive hadron production
accounting for the initial state radiative corrections and the
vacuum polarization corrections (see below).

In the lower energy region, precise measurements of the $R_{\rm uds}(s)$,
which is defined as the $R(s)$ ratio 
for the continuum light hadron (containing u, d and s quarks)
production in $e^+e^-$ annihilation in this paper,
can be used to test 
the validity of the pQCD
calculation~\cite{kuhn_steinhauser,Martin_Outhwaite_Ryskin} in this energy region.
Moreover the ${R_{\rm had}}(s)$ values including the contributions 
from both the continuum hadrons and all
$1^{--}$ resonances at all energies are needed
to calculate the effects of vacuum polarization on the parameters
of the Standard Model, such as
the quantities $\alpha(M^2_Z)$,
the QED running coupling constant
evaluated at the mass of $Z^0$,
and $a_{\mu}=(g-2)/2$, the anomalous magnetic moment 
of the
muon~\cite{davier,martin_plb345_y1995_p558,martin_epjc_31_y2003_p503}.
A large part of uncertainty in the calculation
arises from the uncertainties in the measured inclusive hadronic
cross sections in the open charm threshold region, in which many resonances
overlap. To get credible measurements of the $R_{\rm uds}(s)$ 
for tests of the pQCD calculation and various lowest order cross sections, 
i.e. the measurement of $R_{\rm had}(s)$,
in this energy region,
the overlapping effects have to be clarified clearly.
Near the $D \bar D$ threshold region, say above the $\psi(2S)$ resonance and
below the $D \bar D$ threshold, 
no direct measurement of the $R_{\rm uds}$ is currently available.
The measurements of the $R(s)$~\cite{bes2_2000_2002} in this
region and 
in the region between 2.0 and 3.55 GeV~\cite{bes2_2000_2002},
which were obtained from analyzing the data taken with the BES-II detector
in the period from 1998 to 1999,
can not be used directly to test the
validity, since those 
give the $R_{\rm had}(s)$ rather than
the $R_{\rm uds}$.

In this Letter, we report direct measurements of $R_{\rm uds}$
near the $D \bar D$ threshold region, 
measurements of $R_{\rm uds(c)+\psi(3770)}(s)$ 
above the $D \bar D$ threshold 
and the $R_{\rm had}(s)$ values at 68 energy points
in the region from 3.650 
to 3.872 GeV. 
We here 
define the $R_{\rm uds(c) +\psi(3770)}(s)$ to be the $R(s)$ ratio 
accounting for the contributions from
both the continuum hadron production and the decays for
$\psi(3770)\rightarrow {\rm hadrons}$.
Combining the measurements of the $R_{\rm uds(c)+\psi(3770)}(s)$ and
the cross section for $D\bar D$ production~\cite{xsct_dd_bes}
would give us some information about
non-$D\bar D$ decays of
$\psi(3770)$~\cite{non_dd_decays,r_and_bf,bes_psipp_prmt}
and improve our knowledge about the
nature of $\psi(3770)$.
The data samples used in the analysis were taken
with the BES-II detector~\cite{bes2} at the BEPC Collider in December 2003.

The $\sigma^{\rm obs}_{\rm had}$ is determined by
\begin{equation}
\sigma^{\rm obs}_{\rm had}(s) =\frac{N^{\rm obs}_{\rm had}}
                {L~\epsilon_{\rm had}~\epsilon_{\rm had}^{\rm trig}
                },
\end{equation}
where $N^{\rm obs}_{\rm had}$ is the number of
the observed hadronic events,
$L$ is the integrated luminosity,
$\epsilon_{\rm had}$ is the efficiency for the
detection of inclusive hadronic events 
and $\epsilon_{\rm had}^{\rm trig}$ is the trigger efficiency
for collecting hadronic events in online data acquisition system.

\begin{figure}
\includegraphics[width=8.0cm,height=4.0cm]
{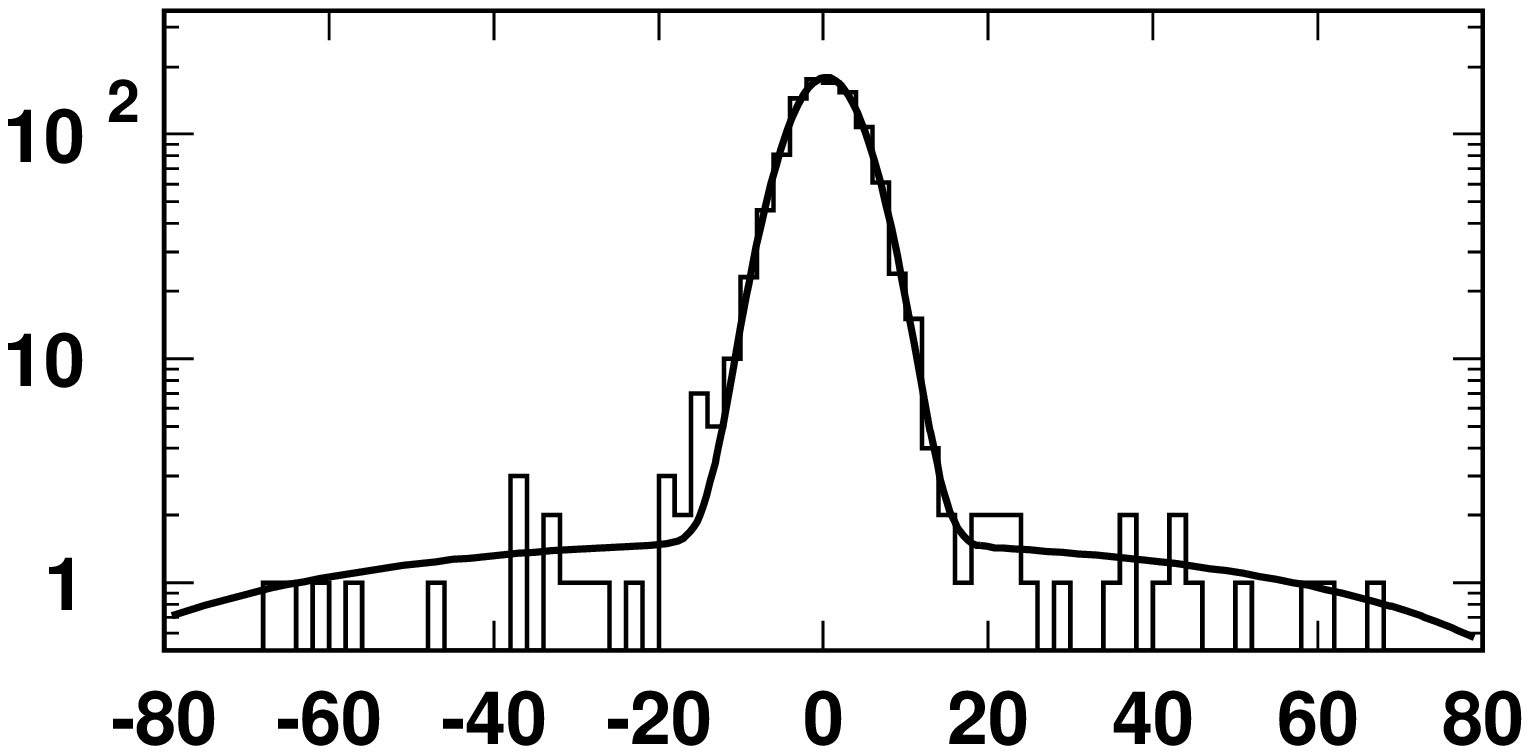}
\put(-115,-5){\bf Z~~[cm]}
\put(-225,25){\rotatebox{90}{\bf No. of Events}}
\caption{The distribution of the event vertex in $z$.}
\label{Zevnt_had}
\end{figure}

The hadronic events are required to have more than 2 good
charged tracks satisfying
selection criteria as described in Ref.~\cite{r_and_bf}. 
To separate some beam-gas associated background events and
cosmic rays background events from hadronic events we calculated the
event vertex in the beam line direction.
Fig.~\ref{Zevnt_had}
shows the distribution of the event vertex of
the accepted events.
Using a Gaussian function to describe the hadronic events plus a second
order polynomial for the background to fit the distribution,
we obtain the number, $N_{\rm had}^{\rm zfit}$, of the candidates for
the hadronic events.
This number of candidates for hadronic events contains
contaminations from
some background sources such as
$e^+e^- \rightarrow \tau^+\tau^-$,
$e^+e^- \rightarrow (\gamma) e^+e^-$,
$e^+e^- \rightarrow (\gamma) \mu^+\mu^-$,
and two-photon processes. The number of
background events, $N_{\rm b}$, due to these processes can be estimated
by means of a Monte Carlo simulation~\cite{r_and_bf}.
Subtracting $N_{\rm b}$ from $N^{\rm zfit}_{\rm had}$ yields
the number of the observed hadronic events, $N^{\rm obs}_{\rm had}$.
The systematic uncertainty in
measuring the produced hadronic events
due to the hadronic event selection criteria
is estimated to be $\sim 2.5\%$~\cite{r_and_bf}.

The integrated luminosities of the data sets are determined
using large-angle Bhabha scattering events
as described in Ref.~\cite{r_and_bf}.
The systematic uncertainty in the measured luminosities
is estimated to be  $\sim 1.9\%$~\cite{r_and_bf}.

\begin{figure}
\includegraphics[width=8.0cm,height=5.5cm]
{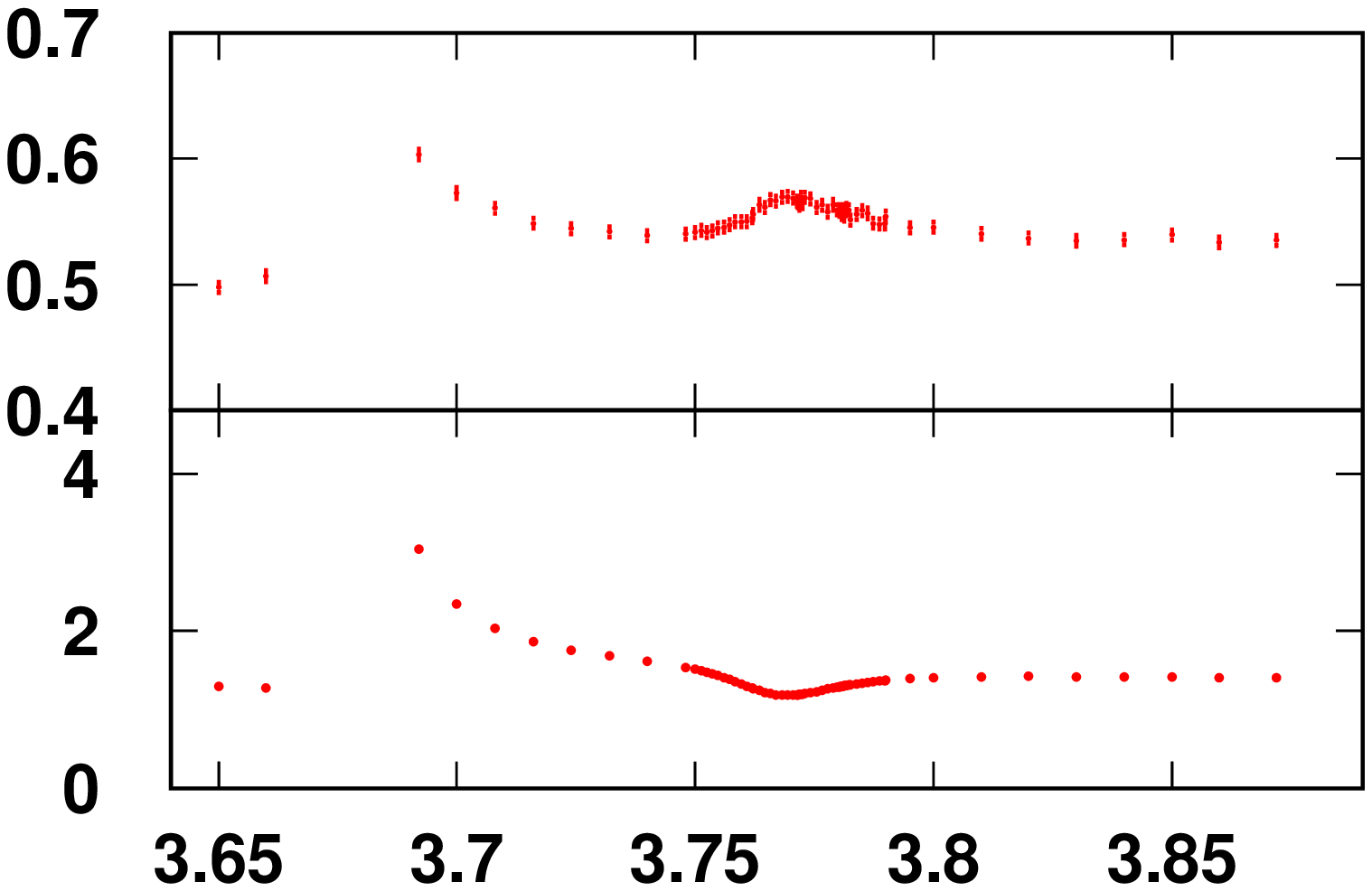}
\put(-130,-7.0){\bf $E_{\rm cm}$ [GeV]}
\put(-225,40){\rotatebox{90}{\bf $(1+\delta)$}}
\put(-225,105){\rotatebox{90}{\bf $\epsilon_{\rm had}$}}
\put(-50,120){\bf (a)}
\put(-50,60){\bf (b)}
\caption{(a) The efficiency versus the nominal c.m. energy;
(b) The ISR factor versus the nominal c.m. energy
(see text).}
\label{xsct_had_ddbar1}
\end{figure}

The detection efficiency for hadronic events is determined
via a special Monte Carlo generator~\cite{zhangdh_gen}
in which the radiative corrections to $\alpha^2$ order
are taken into account.
These generated events are simulated with
the GEANT3-based Monte Carlo simulation.
The systematic uncertainty in the efficiencies due to the generator
is estimated to be 
$\sim 2.0\%$ for reconstruction of the inclusive hadronic events 
from continuum hadrons, $\psi(3770)$ and $\psi(2S)$ decays, 
and to be $\sim 0.7\%$ for reconstruction 
of the continuum hadronic events~\cite{r_and_bf}.
Fig.~\ref{xsct_had_ddbar1}(a) shows the Monte Carlo efficiencies
for the detection of hadronic events
produced at the different nominal c.m. energies, where the error is
statistical.

The trigger efficiencies  are measured to be
$\epsilon_{\rm trig}=(100.0^{+0.0}_{-0.5})\%$
for both the $e^+e^- \rightarrow (\gamma) e^+e^-$ and
$e^+e^-\rightarrow {\rm hadrons}$ events~\cite{r_and_bf}.

\begin{figure}
\includegraphics[width=8.0cm,height=5.5cm]
{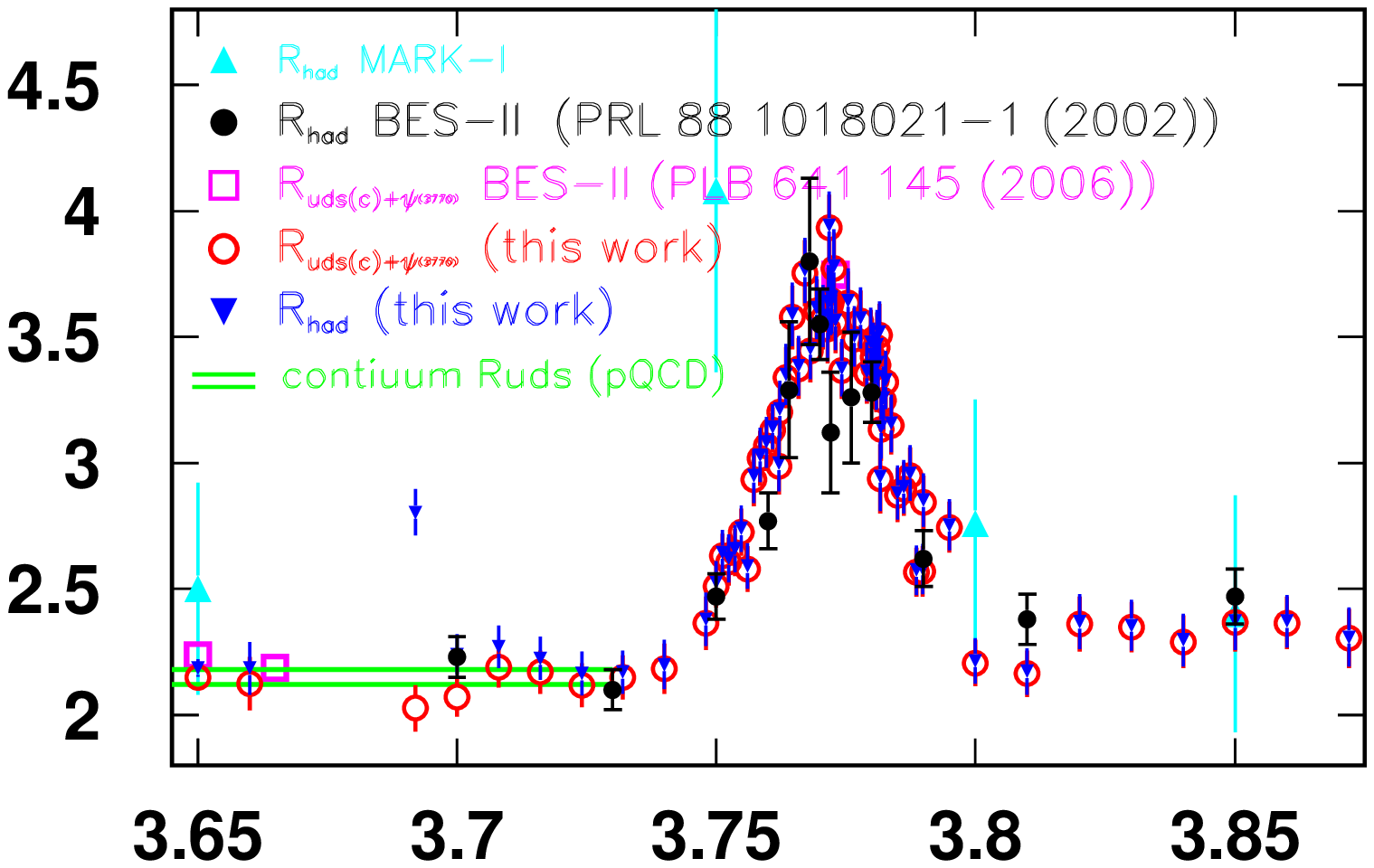}
\put(-130.0,-5.0){\bf{$E_{\rm cm}$ [GeV]}}
\put(-230,20){\rotatebox{90}{\bf $R_{\rm uds(c) +\psi(3770)}(s)$
or $R_{\rm had}(s)$}}  
\caption{The $R_{\rm uds(c) +\psi(3770)}(s)$ or 
$R_{\rm had}(s)$ versus the c.m. energy (see text).
}
\label{r_67pnts}
\end{figure}

To get the 
$\sigma^{0}_{\rm had}(s)$ 
the $\sigma^{\rm obs}_{\rm had}(s)$  
has to be corrected for the 
initial state radiative 
and 
vacuum polarization corrections.
There is no correction for final state radiation needed, since the
$\sigma^{0}_{\rm had}(s)$ includes the final state photon radiation.
The correction factor, $(1+\delta(s))$, is given by
\begin{equation}
(1+\delta(s)) = \frac{\sigma^{\rm exp}(s)}{\sigma^{0}(s)},
\end{equation}
where $\sigma^{\rm exp}(s)$ is the expected cross section
and $\sigma^{0}(s)$ the lowest order cross section
for inclusive hadronic event production.
The $\sigma^{\rm exp}(s)$
can be written as
\begin{equation}
  \sigma^{\rm exp}(s)=\int^{1-\frac {4m_{\pi}^2}{s}}_0  dx \cdot
\frac {\sigma^{0}(s(1-x))}{|1-\Pi(s(1-x))|^2} F(x,s),~~~~~~
\end{equation}
\noindent
where $\sigma^{0}(s(1-x))$ is the total lowest order cross section
in the energy range from 0.28 GeV to $\sqrt{s}$,
$F(x,s)$ is a sampling function~\cite{Kuraev} and
$\frac{1}{|1-\Pi(s(1-x))|^2}$ is the correction factor for the
effects of vacuum polarization
including both the leptonic and hadronic terms
in QED~\cite{r_and_bf,Kuraev}.
Assuming that there are no other new structure and
effects except the $\psi(3770)$ 
and continuum hadrons
in the energy region from 3.729 to 3.872 GeV, we calculate the
radiative correction factors.
In the calculation of the 
cross sections,
the $\psi(3770)$ and $\psi(2S)$
resonance parameters
measured by the BES Collaboration~\cite{bes_psipp_prmt}
are used.
Inserting the resonance parameters of other $1^{--}$
states~\cite{pdg04}
and the $R_{\rm uds}=2.218\pm 0.091$
measured by the BES Collaboration~\cite{r_and_bf} near 
3.65 GeV in Eqs. (6)--(25) of 
Ref.~\cite{r_and_bf} as the initial input,
we obtain the radiative correction factors at the 68 energy points
as shown in Fig.~\ref{xsct_had_ddbar1}(b). 
The uncertainty in $(1+\delta(s))$ is 
less than $1.5\%$~\cite{r_and_bf}. 

The sum of the lowest order cross sections for continuum hadron and $\psi(3770)$
production is given by
\begin{equation}
\sigma^{0}_{\rm uds(c)+\psi(3770)}(s) = 
\frac{\sigma^{\rm obs}_{\rm had}(s)} {(1+\delta(s))} -
\sum_{i}\sigma^{0}_{{\rm Res},i}(s),
\end{equation}
in which $\sum_{i}\sigma^{0}_{{\rm Res},i}(s)$ is the 
contribution from the  $1^{--}$ resonances~\cite{r_and_bf} 
near the c.m. energy of 3.70 GeV 
except $\psi(3770)$.
Eq. (5) indicates 
\begin{equation}
\sigma^0_{\rm uds(c)+\psi(3770)}(s)=
(R_{\rm uds(c)}+R_{\psi(3770)})\cdot \sigma^0_{\mu^+\mu^-}(s),
\end{equation}
\noindent
where 
$R_{\psi(3770)}(s)$ is the $R(s)$ value due to the decays for
$\psi(3770)\rightarrow {\rm hadrons}$.
Dividing the $\sigma^0_{\rm uds(c)+\psi(3770)}(s)$ by 
$\sigma^0_{\mu^+\mu^-}(s)$
yields $R_{\rm uds(c) +\psi(3770)}(s)$ values
as summarized in table~\ref{67pnts_r},
where $R_{\rm uds(c)+\psi(3770)}(s)$ means $R_{\rm uds}(s)$ 
or $R_{\rm uds(c)}(s)+R_{\psi(3770)}(s)$;
below the $D \bar D$ threshold $R_{\rm uds(c)+\psi(3770)}(s)$ 
equals $R_{\rm uds}(s)$; above the
threshold $R_{\rm uds(c)+\psi(3770)}(s)=R_{\rm uds(c)}(s)+R_{\psi(3770)}(s)$
is the $R(s)$ value including both the
continuum hadron production and the decays of 
$\psi(3770)\rightarrow {\rm hadrons}$.
As mentioned above, calculating the effects of
the vacuum polarization on 
$\alpha(M^2_Z)$ and $a^{\rm SM}_{\mu}$
needs the ${R_{\rm had}}(s)$ values,
which can be obtained by dividing 
\begin{equation}
\sigma^{\rm 0}_{\rm had}(s) = 
\frac{\sigma^{\rm obs}_{\rm had}(s)} {(1+\delta(s))}
\end{equation}
by $\sigma^0_{\mu^+\mu^-}(s)$.
Table~\ref{67pnts_r} also lists the ${R_{\rm had}}(s)$ values.
The first errors in the measured $R_{\rm uds(c)+\psi(3770)}(s)$ and $R_{\rm had}(s)$
values listed in the table
are statistical including the point-to-point
systematic uncertainty arising from the statistical uncertainty 
in the measured luminosity
and the uncertainty in the Monte Carlo efficiency,
and the second ones are common systematic,
arising from the 
uncertainties in
luminosity ($\sim 1.9\%$),
in selection of hadronic event ($\sim 2.5\%$),
in Monte Carlo Modeling ($\sim 2.0\%$),
in radiative correction ($\sim 1.5\%$)
for measurements of $R(s)$ values off the $\psi(3770)$ resonance
and in $\psi(3770)$ 
parameters ($\sim 2.7~~\%$)
for those in the $\psi(3770)$ resonance region between 
3.74 to 3.82 GeV.
Adding these uncertainties in quadrature yields the total systematic
uncertainties to be $\sim4.0\%$ and $\sim4.9\%$
for these
outside the $\psi(3770)$ resonance
and within $\psi(3770)$ resonance regions,  
respectively.   
\begin{table*}[t]
\caption{Summary of the $R_{\rm uds(c)+\psi(3770)}(s)$ and $R_{\rm had}(s)$ values 
measured at 68 energy points.
}
\label{67pnts_r}
\begin{center}
\begin{tabular}{c c c c c c } \hline \hline
 c.m. energy  &   $R_{\rm uds(c)+\psi(3770)}(s)$ & $R_{\rm had}(s)$ 
          & c.m. energy  &   $R_{\rm uds(c)+\psi(3770)}(s)$  & $R_{\rm had}(s)$ \\ \hline
  3.6500 & $2.157\pm 0.035\pm 0.086$ & $2.186\pm 0.035\pm 0.087$  &  3.7726 & $3.777\pm 0.145\pm 0.185$ & $3.781\pm 0.145\pm 0.185$ \\
  3.6600 & $2.131\pm 0.105\pm 0.085$ & $2.185\pm 0.105\pm 0.087$  &  3.7730 & $3.563\pm 0.120\pm 0.175$ & $3.567\pm 0.120\pm 0.175$ \\
  3.6920 & $2.034\pm 0.092\pm 0.081$ & $2.803\pm 0.092\pm 0.112$  &  3.7742 & $3.373\pm 0.113\pm 0.165$ & $3.377\pm 0.113\pm 0.165$ \\
  3.7000 & $2.079\pm 0.079\pm 0.083$ & $2.240\pm 0.079\pm 0.089$  &  3.7754 & $3.641\pm 0.125\pm 0.178$ & $3.645\pm 0.125\pm 0.178$ \\
  3.7080 & $2.197\pm 0.083\pm 0.088$ & $2.270\pm 0.083\pm 0.091$  &  3.7766 & $3.498\pm 0.119\pm 0.171$ & $3.502\pm 0.119\pm 0.171$ \\
  3.7160 & $2.177\pm 0.086\pm 0.087$ & $2.224\pm 0.086\pm 0.089$  &  3.7778 & $3.570\pm 0.121\pm 0.175$ & $3.574\pm 0.121\pm 0.175$ \\
  3.7240 & $2.125\pm 0.086\pm 0.085$ & $2.164\pm 0.086\pm 0.086$  &  3.7790 & $3.360\pm 0.117\pm 0.165$ & $3.363\pm 0.117\pm 0.165$ \\
  3.7320 & $2.156\pm 0.086\pm 0.086$ & $2.170\pm 0.086\pm 0.087$  &  3.7798 & $3.477\pm 0.136\pm 0.170$ & $3.480\pm 0.136\pm 0.170$ \\
  3.7400 & $2.190\pm 0.099\pm 0.088$ & $2.200\pm 0.099\pm 0.088$  &  3.7802 & $3.427\pm 0.125\pm 0.168$ & $3.430\pm 0.125\pm 0.168$ \\
  3.7480 & $2.371\pm 0.106\pm 0.116$ & $2.380\pm 0.106\pm 0.116$  &  3.7804 & $3.382\pm 0.137\pm 0.166$ & $3.385\pm 0.137\pm 0.166$ \\
  3.7500 & $2.517\pm 0.085\pm 0.123$ & $2.525\pm 0.085\pm 0.123$  &  3.7808 & $3.336\pm 0.129\pm 0.163$ & $3.340\pm 0.129\pm 0.163$ \\
  3.7512 & $2.637\pm 0.090\pm 0.129$ & $2.644\pm 0.090\pm 0.129$  &  3.7810 & $3.464\pm 0.138\pm 0.170$ & $3.468\pm 0.138\pm 0.170$ \\
  3.7524 & $2.615\pm 0.095\pm 0.128$ & $2.622\pm 0.095\pm 0.128$  &  3.7812 & $3.396\pm 0.130\pm 0.166$ & $3.399\pm 0.130\pm 0.166$ \\
  3.7536 & $2.652\pm 0.093\pm 0.130$ & $2.659\pm 0.093\pm 0.130$  &  3.7814 & $3.514\pm 0.124\pm 0.172$ & $3.518\pm 0.124\pm 0.172$ \\
  3.7548 & $2.733\pm 0.093\pm 0.134$ & $2.739\pm 0.093\pm 0.134$  &  3.7816 & $2.944\pm 0.137\pm 0.144$ & $2.947\pm 0.137\pm 0.144$ \\
  3.7560 & $2.585\pm 0.090\pm 0.127$ & $2.591\pm 0.090\pm 0.127$  &  3.7818 & $3.140\pm 0.125\pm 0.154$ & $3.143\pm 0.125\pm 0.154$ \\
  3.7572 & $2.942\pm 0.107\pm 0.144$ & $2.948\pm 0.107\pm 0.144$  &  3.7822 & $3.253\pm 0.124\pm 0.159$ & $3.257\pm 0.124\pm 0.159$ \\
  3.7584 & $3.025\pm 0.108\pm 0.148$ & $3.031\pm 0.108\pm 0.148$  &  3.7826 & $3.326\pm 0.115\pm 0.163$ & $3.329\pm 0.115\pm 0.163$ \\
  3.7596 & $3.076\pm 0.102\pm 0.151$ & $3.082\pm 0.102\pm 0.151$  &  3.7838 & $3.154\pm 0.114\pm 0.155$ & $3.157\pm 0.114\pm 0.155$ \\
  3.7608 & $3.138\pm 0.089\pm 0.154$ & $3.143\pm 0.089\pm 0.154$  &  3.7850 & $2.879\pm 0.107\pm 0.141$ & $2.882\pm 0.107\pm 0.141$ \\
  3.7620 & $2.992\pm 0.110\pm 0.147$ & $2.998\pm 0.110\pm 0.147$  &  3.7862 & $2.902\pm 0.105\pm 0.142$ & $2.905\pm 0.105\pm 0.142$ \\
  3.7622 & $3.207\pm 0.114\pm 0.157$ & $3.213\pm 0.114\pm 0.157$  &  3.7874 & $2.957\pm 0.111\pm 0.145$ & $2.960\pm 0.111\pm 0.145$ \\
  3.7634 & $3.345\pm 0.122\pm 0.164$ & $3.350\pm 0.122\pm 0.164$  &  3.7886 & $2.571\pm 0.097\pm 0.126$ & $2.574\pm 0.097\pm 0.126$ \\
  3.7646 & $3.585\pm 0.126\pm 0.176$ & $3.590\pm 0.126\pm 0.176$  &  3.7898 & $2.576\pm 0.099\pm 0.126$ & $2.579\pm 0.099\pm 0.126$ \\
  3.7658 & $3.381\pm 0.119\pm 0.166$ & $3.386\pm 0.119\pm 0.166$  &  3.7900 & $2.849\pm 0.106\pm 0.140$ & $2.852\pm 0.106\pm 0.140$ \\
  3.7670 & $3.760\pm 0.130\pm 0.184$ & $3.764\pm 0.130\pm 0.184$  &  3.7950 & $2.751\pm 0.101\pm 0.135$ & $2.754\pm 0.101\pm 0.135$ \\
  3.7682 & $3.451\pm 0.124\pm 0.169$ & $3.455\pm 0.124\pm 0.169$  &  3.8000 & $2.212\pm 0.091\pm 0.108$ & $2.215\pm 0.091\pm 0.108$ \\
  3.7694 & $3.611\pm 0.125\pm 0.177$ & $3.615\pm 0.125\pm 0.177$  &  3.8100 & $2.171\pm 0.092\pm 0.087$ & $2.173\pm 0.092\pm 0.087$ \\
  3.7706 & $3.580\pm 0.123\pm 0.175$ & $3.584\pm 0.123\pm 0.175$  &  3.8200 & $2.367\pm 0.109\pm 0.095$ & $2.369\pm 0.109\pm 0.095$ \\
  3.7714 & $3.538\pm 0.139\pm 0.173$ & $3.543\pm 0.139\pm 0.173$  &  3.8300 & $2.354\pm 0.101\pm 0.094$ & $2.355\pm 0.101\pm 0.094$ \\
  3.7716 & $3.634\pm 0.146\pm 0.178$ & $3.638\pm 0.146\pm 0.178$  &  3.8400 & $2.296\pm 0.104\pm 0.092$ & $2.297\pm 0.104\pm 0.092$ \\
  3.7718 & $3.939\pm 0.133\pm 0.193$ & $3.943\pm 0.133\pm 0.193$  &  3.8500 & $2.372\pm 0.115\pm 0.095$ & $2.373\pm 0.115\pm 0.095$ \\
  3.7720 & $3.636\pm 0.134\pm 0.178$ & $3.640\pm 0.134\pm 0.178$  &  3.8600 & $2.371\pm 0.105\pm 0.095$ & $2.372\pm 0.105\pm 0.095$ \\
  3.7722 & $3.652\pm 0.143\pm 0.179$ & $3.656\pm 0.143\pm 0.179$  &  3.8720 & $2.308\pm 0.117\pm 0.092$ & $2.309\pm 0.117\pm 0.092$ \\
\hline \hline
\end{tabular}
\end{center}
\end{table*}

Weighting the first 8 $R_{\rm uds(c)+\psi(3770)}(s)$ values 
below the $D \bar D$ threshold with the statistical error 
yields 
$$R_{\rm uds}=2.141 \pm 0.025 \pm 0.085,$$
\noindent
which can directly be compared to
$R^{\rm pQCD}_{\rm uds}=2.15 \pm 0.03$
calculated  by pQCD~\cite{martin_plb345_y1995_p558,harlander_steinhauser}.

Figure~\ref{r_67pnts} displays the $R_{\rm uds(c)+\psi(3770)}(s)$ 
and/or $R_{\rm had}(s)$ values from this measurements, together
with those measured by the MARK-I Collaboration~\cite{mark1},
by the BES Collaboration
from analysis of different data
samples~\cite{bes2_2000_2002}
and the continuum $R^{\rm pQCD}_{\rm uds}$ 
calculated  by pQCD~\cite{harlander_steinhauser,martin_plb345_y1995_p558}, which is
shown by two straight lines indicating $\pm 1\sigma$ error interval.
There is one thing which we would like to point out. 
In the measurements of the $R(s)$ values, 
the measurements reported in Ref.~\cite{r_and_bf} 
are the quantity 
$R_{\rm uds(c)+\psi(3770)}(s)$
which equals $R_{\rm uds}$ below the $D \bar D$ threshold
and equals $R_{\rm uds(c)}(s)+R_{\psi(3770)}(s)$
above the $D \bar D$ threshold;
while the measurements reported from this work are the quantities
$R_{\rm uds}$,
$R_{\rm uds(c)+\psi(3770)}(s)$ and
$R_{\rm had}(s)$,
the other measurements are only the 
quantity
$R_{\rm had}(s)$.

In summary, we measured the $R_{\rm uds(c)+\psi(3770)}(s)$ 
and $R_{\rm had}(s)$ values 
in the energy region from 3.650 to 3.872 GeV
with the systematic uncertainty of 
$\sim4.0\%$ for the continuum $R_{\rm uds}$ and the $R_{\rm had}(s)$
outside the $\psi(3770)$ resonance
regions, 
and with the uncertainty of
$\sim4.9\%$ for the $R_{\rm uds(c)+\psi(3770)}(s)$ 
and the $R_{\rm had}(s)$ in the $\psi(3770)$ resonance region.
These are improved measurements with respect to 
the earlier measurements~\cite{bes2_2000_2002,mark1}
in both the accuracy and the number of energy points.
These improved measurements of the $R_{\rm had}(s)$ 
are expected to provide an improvement in the precision of the calculated
values of $\alpha(M^2_Z)$~\cite{b_pietrzyk} and 
$a^{\rm SM}_{\mu}$~\cite{b_pietrzyk,m_davier_y2003,martin_epjc_31_y2003_p503}.
We obtained the 
continuum $R_{\rm uds}$
near the $D\bar D$ threshold region to be 
$R_{\rm uds}=2.141 \pm 0.025 \pm 0.085$,
which is in excellent agreement with 
$R^{\rm pQCD}_{\rm uds}=2.15\pm0.03$~\cite{harlander_steinhauser,martin_plb345_y1995_p558}
predicted by pQCD in this energy region.
This measured $R_{\rm uds}$ can directly be used to evaluate the strong
coupling constant 
$\alpha_s(s)$ at the energy scale of $\sim 3$ GeV.
In this analysis, we clarified the overlapping effects of
different processes to the $R_{\rm had}(s)$.
This is important for test of the validity of the pQCD calculation
near the $D\bar D$ threshold region and for correctly using the data in
calculation of $\alpha(M^2_Z)$ and $a^{\rm SM}_{\mu}=(g-2)/2$ with the help of the
pQCD prediction.
\vspace{0.05mm}

   The BES collaboration thanks the staff of BEPC for their hard efforts.
This work is supported in part by the National Natural Science Foundation
of China under contracts
Nos. 19991480,10225524,10225525, the Chinese Academy
of Sciences under contract No. KJ 95T-03, the 100 Talents Program of CAS
under Contract Nos. U-11, U-24, U-25, and the Knowledge Innovation Project
of CAS under Contract Nos. U-602, U-34(IHEP); by the
National Natural Science
Foundation of China under Contract No.10175060(USTC),and
No.10225522(Tsinghua University).

\end{document}